\documentclass[prd,superscriptaddress,amsfonts,amssymb,amsmath,twocolumn,showpacs]{revtex4-2}
\usepackage{bm}
\usepackage{amsfonts}
\usepackage{latexsym}
\usepackage[latin1]{inputenc}
\usepackage{graphicx}
\usepackage{amsmath}
\usepackage{palatino}
\usepackage{mathpazo}
\usepackage{textcomp}
\usepackage{float}
\usepackage{booktabs}
\usepackage{dcolumn}
\usepackage{booktabs}
\usepackage{multirow}
\usepackage{hyperref}
\hypersetup{colorlinks,citecolor=red}
\usepackage{amsmath}
\usepackage{xcolor}
\usepackage{orcidlink}
\usepackage[caption=false]{subfig}
\usepackage{commath}

\def\jnl@style{\it}
\def\aaref@jnl#1{{\jnl@style#1}}

\def\aaref@jnl#1{{\jnl@style#1}}

\def\aj{\aaref@jnl{AJ}}                   
\def\apj{\aaref@jnl{ApJ}}                 
\def\apjl{\aaref@jnl{ApJ}}                
\def\apjs{\aaref@jnl{ApJS}}               
\def\apss{\aaref@jnl{Ap\&SS}}             
\def\aap{\aaref@jnl{A\&A}}                
\def\aapr{\aaref@jnl{A\&A~Rev.}}          
\def\aaps{\aaref@jnl{A\&AS}}              
\def\mnras{\aaref@jnl{Mon.~Not.~Roy.~Astron.~Soc.}}             
\def\prd{\aaref@jnl{Phys.~Rev.~D}}        
\def\prc{\aaref@jnl{Phys.~Rev.~C}}  
\def\prl{\aaref@jnl{Phys.~Rev.~Lett.}}    
\def\qjras{\aaref@jnl{QJRAS}}             
\def\skytel{\aaref@jnl{S\&T}}             
\def\ssr{\aaref@jnl{Space~Sci.~Rev.}}     
\def\zap{\aaref@jnl{ZAp}}                 
\def\nat{\aaref@jnl{Nature}}              
\def\aplett{\aaref@jnl{Astrophys.~Lett.}} 
\def\apspr{\aaref@jnl{Astrophys.~Space~Phys.~Res.}} 
\def\physrep{\aaref@jnl{Phys.~Rep.}}      
\def\physscr{\aaref@jnl{Phys.~Scr}}       
\def\commat{\aaref@jnl{Comm.~Math.~Phys.}}              
\def\science{\aaref@jnl{Science}}               
\def\cqg{\aaref@jnl{Classical Quant.~Grav.}}            
\def\jpcs{\aaref@jnl{JPCS}}                                     
\def\ijmpd{\aaref@jnl{Int.~J.~Mod.~Phys.~D}}                    
\def\grg{\aaref@jnl{Gen.~Relat.~Gravit.}}               
\def\rpp{\aaref@jnl{Rep.~Prog.~Phys.}}          
\def\npa{\aaref@jnl{Nucl.~Phys.~A}}        
\def\lrr{\aaref@jnl{Living Rev.~Rel.}}                   
\def\jcap{\aaref@jnl{J.~Cosmology Astropart.~Phys.}}    
\def\rmp{\aaref@jnl{Rev.~Mod.~Phys.}}   
\def\epjc{\aaref@jnl{Eur.~Phys.~J.~C}}

\newcommand{\udt}[3]{#1^{#2}_{\phantom{#2}#3}}

\allowdisplaybreaks[1]

\begin{document}

\color{black}       
\title{\bf Scalar field induced dynamical evolution in teleparallel gravity}

\author{B. Mishra\orcidlink{0000-0001-5527-3565} }
\email{bivu@hyderabad.bits-pilani.ac.in }
\affiliation{Department of Mathematics, Birla Institute of Technology and Science-Pilani, Hyderabad Campus, Hyderabad-500078, India.}

\author{S. A. Kadam\orcidlink{0000-0002-2799-7870}}
\email{k.siddheshwar47@gmail.com}
\affiliation{Department of Mathematics, Birla Institute of Technology and Science-Pilani, Hyderabad Campus, Hyderabad-500078, India}

\author{S. K. Tripathy \orcidlink{0000-0001-5154-2297}}
\email{tripathy\_sunil@rediffmail.com}
\affiliation{Department of Physics, Indira Gandhi Institute of Technology, Sarang, Dhenkanal, Odisha-759146, India.}

\begin{abstract}
In this paper, we investigate the role of scalar field potentials in the dynamical evolution of the Universe. A gravity theory with a non-minimally coupled scalar field with torsion in the geometrical action simulating effective dark energy is considered to study an extended matter bounce scenario. The dynamical behaviour of the equation of state parameter has been studied near the bouncing epoch. Keeping in mind the inflationary behaviour near the bounce, five different scalar field potential functions are explored, and their effect on the equation of state parameter is investigated.
\end{abstract}

\maketitle

\section{Introduction} \label{SEC I}
Teleparallel gravity (TG) is another geometric theory in which the Levi-Civita connection is replaced with the teleparallel Weizenb$\ddot{o}$ck connection \cite{Weitzenbock1923}. It expresses the geometric deformations through torsion rather than curvature, and all measures of curvature vanish identically. However, the regular curvature form does not vanish. Both the connections can be related, and the torsion scalar can be produced, which is equal to the regular Ricci scalar up to some boundary term. Therefore, the field equations of TG are also dynamically equivalent to General Relativity(GR)  \cite{Bahamonde:2021gfp,Aldrovandi:2013wha,Weitzenbock1923,Cai:2015emx,Krssak:2018ywd,Kadam:2022lgq}. The division of the torsion scalar and boundary term resulted in a weaker Lovelock theorem \cite{Lovelock:1971yv,Gonzalez:2015sha}. This will produce more general theories of second order
such as $f(T)$ gravity \cite{Bahamonde:2019shr,Krssak:2018ywd}. Some of the works in the formulation of $f(T)$ theories of gravity can be found in literature \cite{Ferraro:2008ey,Linder:2010py,Chen:2010va,PhysRevD.94.023525,Duchaniya:2022rqu}. The solar system test and its constraints have been studied in Ref. \cite{Iorio:2012cm,Farrugia:2016xcw}. The stability of flat FLRW metric in $f(T)$ gravity by analyzing the small perturbations, $\delta$ about the Hubble parameter and the matter-energy density ($\delta_{m}$) has been studied in Ref. \cite{Farrugia:2016qqe}. The galactic rotation dynamics \cite{Finch:2018gkh}; probing $f(T)$ gravity with gravitational time advancement \cite{Deng:2018ncg} are also shown. The Noether symmetry approach admits a matter-dominated solution for a special case of a power law with the associated conserved current \cite{SK2017100}. Different epochs of the Universe using dynamical system analysis are studied using the $f(T)$ gravity formalism \cite{Duchaniya:2022rqu,DUCHANIYA2024f(T)}. Moreover, this gravity solves the particle horizon problem in a
spatially flat Universe \cite{Ferraro:2006jd}.

The $f(T,\phi)$ gravity, an extension of $f(T)$ gravity, is the generalised scalar torsion gravity, where $\phi$ is the canonical scalar field. In the gravitational action of $f(T,\phi)$ gravity, the scalar field is non-minimally
connected with torsion scalar \cite{Xu:2012jf}. In this gravity theory, Duchaniya et al. \cite{Duchaniya_2023} have investigated the evolutionary history of the Universe using the dynamical system analysis. Espinoza and Otalora \cite{Gonzalez_Espinoza_2020A}  have added the kinetic term in $f(T,\phi)$ gravity and developed the primordial density perturbation. Moreover, there are some investigations available in $f(T,\phi)$ gravity theory, such as the scalar perturbation in Ref.\cite{Gonzalez-Espinoza:2021mwr} and cosmological dynamics of dark energy in Ref. \cite{Gonzalez-Espinoza:2020jss}. Trivedi et al. \cite{trivedi2023cosmological} have shown the existence of type IV singularity in a limited range of initial conditions in $f(T,\phi)$ gravity. The reconstruction method for determining both the non-minimal coupling function and the scalar potential through the parametrization (or attractor) of the scalar spectral index has been performed in Ref.\cite{Gonzalezreconstruction2021}.

The inflationary paradigm proposed to address some long standing issues in cosmology seems to be incomplete due to the big bang singularity. During the big bang, there is a complete breakdown of physics. It is quite natural to think beyond the big bang singularity, probably with a bounce replacing the big bang. In Loop Quantum Gravity (LQG), quantum effects at the Planck scale are able to produce a bounce. The repulsive quantum geometrical effects replace the big bang singularity by a quantum bounce. Different inflationary potentials have been suggested in the literature, which may have some bearings on the bounce occurring at some phases of cosmic evolution. In this study, we wish to consider an $f(T,\phi)$ gravity theory with the scalar field interacting with inflationary potentials to study their roles on the cosmic behaviour during the bounce.

The paper is organised as follows: in Sec.~\ref{SECII}, we present the basic equations of  $f(T,\phi)$ gravity. In the $f(T,\phi)$ gravity, the action contains the contribution from a scalar field along with an interaction potential. In Sec.~\ref{SECIII}, the dynamical evolution of the Universe in the framework of the said gravity theory is studied where we propose a specific form of the functional $f(T,\phi)$. The cosmic dynamics for different inflationary scalar field potentials are obtained numerically in Sec.~\ref{SECIV}. In Sec.~\ref{SECV}, a brief summary and conclusion of the work are presented.

\section{Field Equations of \texorpdfstring{$f(T,\phi)$}{} gravity }\label{SECII}

The action of $f(T,\phi)$ gravity\cite{Gonzalez-Espinoza:2021mwr} is,
\begin{equation}\label{1}
    S =\int d^{4}x~~e[X+f(T,\phi)]+S_{m}+S_{r}\,,
\end{equation}
where $X= -\partial_{\mu} \phi \partial^{\mu} \phi/2$ be the kinetic energy. The matter action and radiation action are respectively denoted as $S_{m}$ and $S_{r}$. The torsion $T$ is given by
\begin{equation}
T = S_{\theta}^{\mu \nu} T_{\mu \nu}^{\theta},
\end{equation}
where $S_{\theta}^{\mu \nu}$ be the superpotential and $T_{\mu \nu}^{\theta}$ be the torsion tensor. The determinant of the tetrad field $e^{A}_{\mu}$ with $A = 0,1,2,3$ becomes  $e = \det[e^A_{\mu}] = \sqrt{-g}$. The tetrad field satisfies the orthogonality condition, 
\begin{equation}
e^{\mu}_Ae^B_{\mu}=\delta_A^B,
\end{equation}
and connects the metric tensor $g_{\mu \nu}$ and Minkowski tangent space metric $\eta_{AB}=diag(-1,1,1,1)$  through the relation 
\begin{equation}
g_{\mu \nu}=\eta_{AB} e_{\mu}^{A} e_{\nu}^{B}.    
\end{equation}

The torsion tensor is obtained from the spin connection $\udt{\omega}{A}{B\mu}$ and the contortion tensor $    K^{\mu \nu}_{~~~\theta}\equiv \frac{1}{2}(T^{\nu \mu}_{~~~\theta}+T_{\theta}^{~~\mu \nu}-T^{\mu \nu}_{~~~\theta})$ as 
\begin{equation}\label{2}
T_{\mu \nu}^{\theta} = e^{\theta}_{A}\partial_{\mu} e^{A}_{\nu}-e^{\theta}_{A}\partial_{\nu}e^{A}_{\mu}+e^{\theta}_{A} \omega^{A}_{B\mu}e^{B}_{\nu}-e^{\theta}_{A} \omega^{A}_{B\nu}e^{B}_{\mu}.
\end{equation}
The superpotential is expressed in terms of contortion tensor as,
\begin{equation} \label{4}
    S_{\theta}^{~~\mu \nu}\equiv\frac{1}{2}(K^{\mu \nu}_{~~~\theta}+\delta^{\mu}_{\theta}T^{\alpha \nu}_{~~~\alpha}-\delta^{\nu}_{\theta}T^{\alpha \mu}_{~~~\alpha})\,.
\end{equation} 
Along with this, the spin connection becomes zero because of the Weitzenb$\ddot{o}$ck gauge. A variation of the action with the tetrad field yields the field equations of $f(T,\phi)$ gravity  as\cite{Hohmann:2018rwf},
\begin{equation}\label{5}
T = -R + {2}e^{-1}\partial_{\mu}(eT^{\alpha\mu}_{~~\alpha})\,.
\end{equation}
Further for the homogeneous and isotropic flat FLRW space-time,
\begin{equation}
ds^{2}=-dt^{2}+a^{2}(t)[dx^2+dy^2+dz^2],   
\end{equation}
we obtain the field equations [eq. \eqref{5}] as,
\begin{eqnarray}
f(T,\phi)-X-2Tf_{T}&=&\rho_{m}+\rho_{r}, \label{7}\\
f(T,\phi)+X-2Tf_{T}-4\dot{H}f_{T}-4H\dot{f}_{T} &=& -p_{r},\label{8}\\
-3H\dot{\phi}-\ddot{\phi}+f_{\phi}&=& 0. \label{9}
\end{eqnarray}
The tetrad becomes $e^{A}_{\mu} = diag(1,a,a,a)$, where $a(t)$ is the scale factor and $H\equiv\frac{\dot{a}}{a}$ is the Hubble parameter. The over-dot represents the derivative with respect to the cosmic time $t$. The energy density comprises the matter and radiation energy densities respectively denoted as $\rho_{m}$, $\rho_{r}$. The matter field is assumed to be composed of pressureless dust along with the radiation pressure $p_{r}=\frac{1}{3} \rho_r$. Also, the torsion scalar is $T=6H^{2}$. For a non-minimal coupling function $f(T,\phi)$ with a scalar field potential $V(\phi)$, we may have the choice
\begin{equation}\label{10}
f(T,\phi)=-\left(\frac{T}{2}+G(T)\right)-V(\phi)\,,
\end{equation}
where $G(T)$ be an arbitrary function of torsion scalar $T$. For the sake of representation, we denote $f\equiv f(T,\phi)$ such that, $f_{T}=\frac{\partial f}{\partial T}$ and $f_{\phi}=\frac{\partial f}{\partial \phi}$.

The equivalent Friedmann equations of the $f(T,\phi)$ cosmology are written as
\begin{eqnarray}
   & 3H^2=\frac{\dot{\phi}^2}{2}+V(\phi)-2TG_{T}+G(T)+\rho_{m}+\rho_{r} ,\label{11}\\
   &2\dot{H}+3H^2 = -\Big[\frac{\dot{\phi}^2}{2}-V(\phi)+2TG_{T}-G(T)\nonumber\\
   & ~~~~~~~~~~~~~~~~~~~~~~~~~~~~~~~~~+4\dot{H}(G_{T}+2TG_{TT})-\frac{1}{3}\rho_{r}\Big] ,\label{12}\\
   &0= \ddot{\phi}+3H\dot{\phi}+V_{\phi}(\phi)~~~~~~~~~~~.\label{13}
\end{eqnarray}
The subscripts $T$ and $\phi$ in a field variable denote the partial derivatives with respect to the torsion scalar $T$ and scalar field $\phi$.\\
The Friedmann equations for a cosmic fluid comprising dust matter and radiation are given as
\begin{eqnarray}
3H^{2} &=& \rho_{m}+\rho_{r}+\rho_{de}, \label{14}\\
2\dot{H}+3H^2 &=& -(p_r+p_{de}), \label{15}
\end{eqnarray}
Hence we obtain the dark energy density and dark energy pressure in $f(T,\phi)$ gravity as
\begin{align}
    \rho_{de} &= \frac{\dot{\phi}^2}{2}+V(\phi)-2TG_{T}+G(T), \label{16}\\
    p_{de} &= \frac{\dot{\phi}^2}{2}-V(\phi)+2TG_{T}-G(T)+4\dot{H}(G_{T}+2TG_{TT}). \label{17}
\end{align}

Subsequently, the equation of state (EoS) parameter for the dark energy phase becomes,
 \begin{equation}
    \omega_{de}=-1+\frac{\dot{\phi}^2+4\dot{H}(G_{T}+2TG_{TT})}{\frac{\dot{\phi}^2}{2}+V(\phi)-2TG_{T}+G(T)}.
\end{equation}
One may note that, for a given bouncing scenario, the behaviour of the EoS parameter is governed by the scalar field potential and the contribution coming from the geometrically modified gravity theory. We consider here a specific action of the geometrically modified gravity theory, and in that framework, different potentials are applied to understand the evolutionary behaviour of the EoS parameter, particularly near the bounce.

\section{Dynamical evolution of the Universe in $f(T,\phi)$ gravity}\label{SECIII}
In this section, we present the general formulation of the dynamical features of the Universe within the framework of $f(T,\phi)$ gravity theory. An extended matter bounce scenario is analysed in this framework. The dynamical behaviour of the Universe depends on the choice of the functional form of the non-minimal coupling function $f(T,\phi)$. We may take a function as \cite{Cai:2015emx},
\begin{equation}
    G(T) = \beta~T\left(1+T^{n-1}\right),
\end{equation}
where $\beta$ and $n$ are model parameters; the model parameters are fixed from the behaviour of the model under bouncing dynamics. 

The energy density and the pressure for the effective dark energy become
\begin{eqnarray}
    \rho_{de} &=& \frac{\dot{\phi}^2}{2}+V(\phi)-2\beta~T\left(1+nT^{n-1}\right),\\
    p_{de} &=& \frac{\dot{\phi}^2}{2}-V(\phi)+2\beta~T\left(1+nT^{n-1}\right)\nonumber\\
        &+&4\beta\dot{H}\left(1-n+2n^2\right)T^{n-1}.
\end{eqnarray}

For the effective dark energy density to be positive, we have the condition for the kinetic energy density of the scalar field
\begin{equation}
   \frac{\dot{\phi}^2}{2} \geq 2\alpha~T\left(1+nT^{n-1}\right)-V(\phi).
\end{equation}

The extended matter bounce scenario is given by
\begin{equation}
    H=\frac{4\gamma~t}{3(1+\gamma t^2)},
\end{equation}
with $\dot{H}=\frac{4\gamma(1-\gamma t^2)}{3(1+\gamma t^2)^2}$. 
Here $\gamma$ is a parameter that characterizes the bouncing behaviour of the Universe, and $t_B=0$ is the bouncing epoch. The occurrence of a cosmic bounce is believed to be affected by the choice of the scalar field potential. Also, the scalar field potential plays a vital role in providing an explanation to cosmic dynamics near the bounce. For a given bouncing scenario, the scalar field may be obtained from the Klein-Gordon equation
\begin{equation}\label{18}
    \ddot{\phi}+3H\dot{\phi}+V_{\phi}(\phi)=0,
\end{equation}
by incorporating a potential. 

We wish to consider some of the well-known potentials initially suggested for the inflationary scenario and investigate their role on the cosmic dynamics during the bouncing epoch. Since, for some of the chosen potentials, obtaining the scalar field for the extended matter bounce is not straightforward and analytical, we adopt the numerical integration procedure to get a qualitative behaviour of the cosmic dynamics.

\section{Effect of scalar field potential in cosmic dynamics and numerical solution}\label{SECIV}
In the framework of $f(T,\phi)$, different dynamical quantities depend on the scalar field and its interaction. We consider below some of the well known scalar field potentials to obtain the evolutionary aspect.
\subsection{Non-interacting scalar field}
Let us assume that, the scalar field is non-interacting and only has the kinetic contribution to the Lagrangian. For such a case, the Klein-Gordon equation becomes
\begin{equation}
    \frac{1}{\dot{\phi}^2}~d\dot{\phi}^2 = -6H~dt,
\end{equation}
which on integration provides
\begin{equation}
    \dot{\phi} =\frac{k}{(1+\gamma t^2)^2},\label{scalarfield0}
\end{equation}
where $k$ is a positive integration constant and may be chosen as unity for brevity. The corresponding dark energy density and dark energy pressure becomes
\begin{eqnarray}
    \rho_{de} &=& \frac{1}{2(1+\gamma t^2)^4}-2\beta~T\left(1+nT^{n-1}\right),\\
    p_{de} &=& \frac{1}{2(1+\gamma t^2)^4}+2\beta~T\left(1+nT^{n-1}\right)\nonumber\\
       &+&4\beta\dot{H}\left(1-n+2n^2\right)T^{n-1}.
\end{eqnarray}
At bounce, $t_B=0$ and we have vanishing Hubble parameter and torsion. Therefore, at the bounce, the energy density and pressure of the effective dark energy become $\rho_{de}=p_{de}=\frac{1}{2}.$ Another interesting aspect is that the sum of the dark energy density and dark energy pressure becomes 1 which shows that the null energy condition for such a cosmic scenario with non-interacting scalar fields is not violated at least at the bounce.

\subsection{Scalar field with a constant potential}
Let there prevail a constant interaction potential among the scalar field such that $V(\phi)=V_0=\text{constant}>0$. In this case, also we obtain the scalar field as that in \eqref{scalarfield0} and we have 
\begin{eqnarray}
    \rho_{de} &=& \frac{1}{2(1+\gamma t^2)^4}+V_0-2\beta~T\left(1+nT^{n-1}\right),\\
    p_{de} &=& \frac{1}{2(1+\gamma t^2)^4}-V_0+2\beta~T\left(1+nT^{n-1}\right)\nonumber\\
    &+&4\beta\dot{H}\left(1-n+2n^2\right)T^{n-1}.
\end{eqnarray}

At the bounce, we have now $\rho_{de}=\frac{1}{2}+V_0$ and $p_{de}=\frac{1}{2}-V_0$ so that EoS parameter at bounce becomes $\omega_B=\frac{1-2V_0}{1+2V_0}$. The value of $\omega_B$ obviously depends on the choice of the repulsive scalar field potential.

\subsection{Scalar field interacting through a chaotic potential}
The chaotic potential is given by \cite{LINDE1983177}
\begin{equation}
    V(\phi)=\frac{1}{2}m^2\phi^2,
\end{equation}
where $m=1.26\times 10^{-6}~m_{pl}$. This chaotic potential may be used in the Klein-Gordon equation to obtain the scalar field. However, the integration is to be carried out numerically which requires an initial condition for the scalar field at least during the bounce. In view of this, we are required to choose $\phi_B$. It is certain that different initial values of the scalar field will lead to different numerical values of the dynamical parameters. However, we have checked that, a change in the initial value of $\phi_B$ by a small amount does not change the general dynamical behaviour of the EoS parameter and other dynamical quantities. In view of this, we consider only the specific case $\phi_B=1$.

\subsection{Scalar field with a generalized Starobinsky potential}
The Starobinsky model in $f(R)$ theory involves the use of a functional form $f(R)=R+\alpha R^2$ in the Einstein-Hilbert action which is equivalent to a scalar field model with a potential. For E-models, a generalized version of this potential is used \cite{PhysRevD.88.085038}:
   \begin{equation}
    V(\phi) =V_0 \left[1-exp\left(-\frac{\phi}{\sqrt{6} \alpha}\right)\right]^2,
\end{equation}
where $V_0$ is the strength of the potential and $\alpha$ is a dimensionless  positive constant.
\subsection{Scalar field with $\alpha$-attractor potential}
We consider here an $\alpha$-attractor potential in the form of the generalized T-model potential of superconformal inflationary models given by \cite{saeed2024universal}
\begin{equation}
V(\phi)= V_0~ \tanh^2{\frac{\phi}{\sqrt{6}\alpha}}.
\end{equation}

\begin{figure}[H]
    \centering
    \includegraphics[width=72mm]{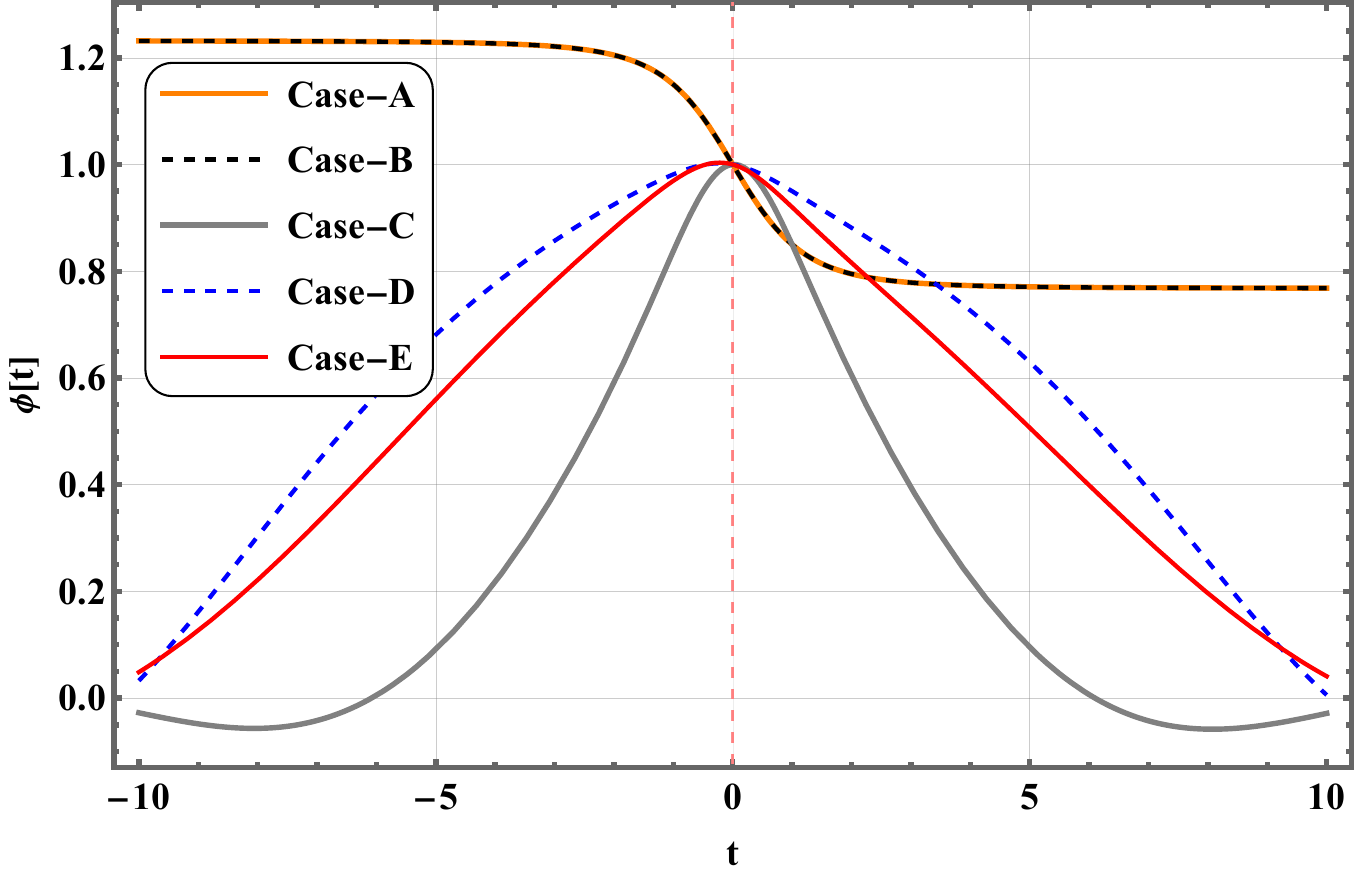}
    \caption{Evolution of scalar field $\phi(t)$ in cosmic time $t$ for the parametric values $\beta=-0.3, n=2.5, \gamma=0.4$. } \label{phi_t} \label{fig1}
\end{figure}

In FIG.-\ref{fig1}, the scalar field as obtained using different scalar field potentials is shown. The scalar field without an interaction potential or with a constant potential behaves in the same manner and are derived analytically. However, for all other cases of potential, we obtain the scalar field numerically assuming an initial value $\phi_B=1$. More or less, the potentials meant to describe the inflationary scenario lead to a universal behaviour for the scalar field.

\begin{figure}[H]
    \centering
    \includegraphics[width=72mm]{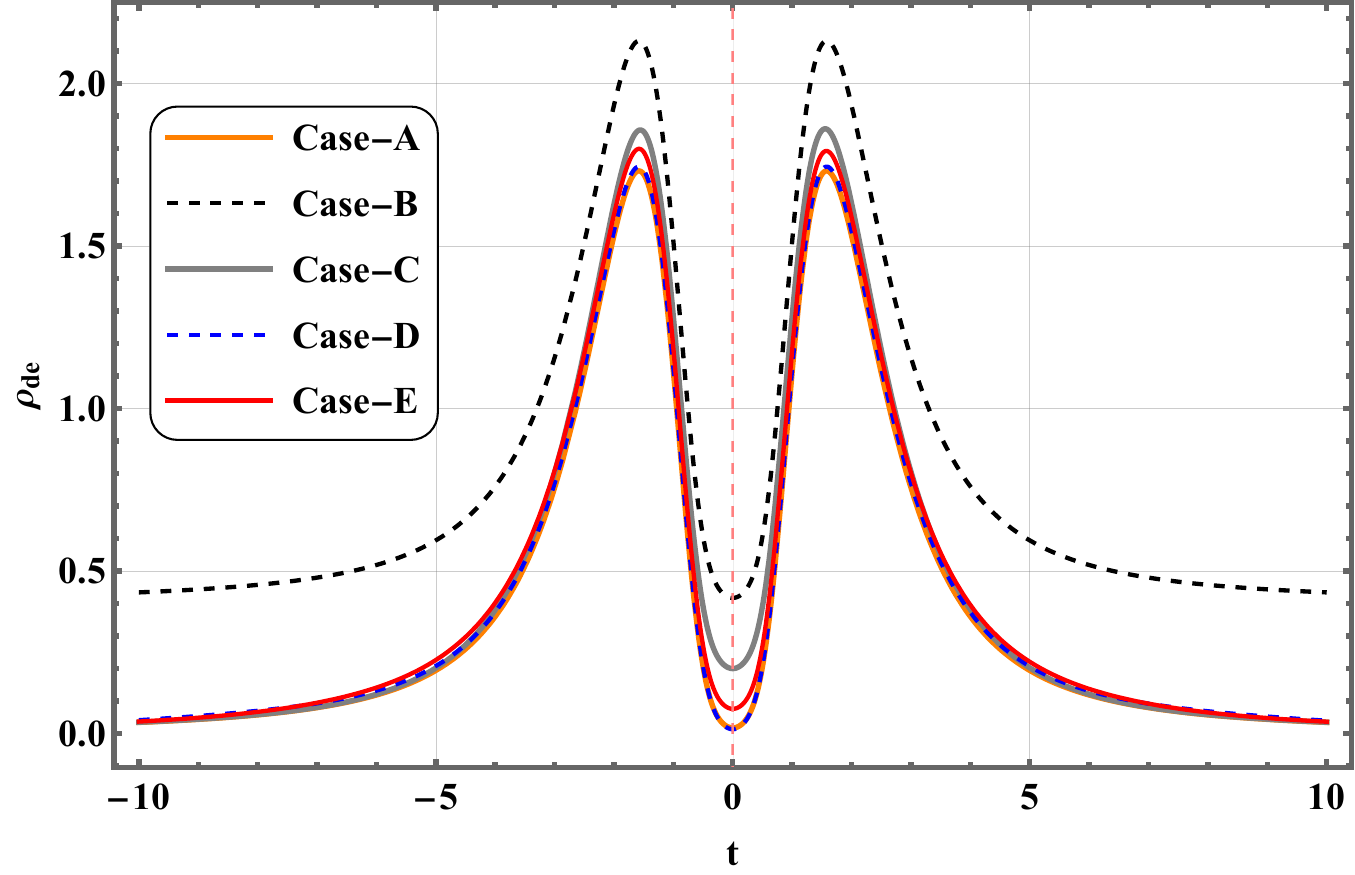}
    \caption{ Evolution of effective energy density ($\rho_{de}$)  in cosmic time $t$ for the parametric values $\beta=-0.3, n=2.5, \gamma=0.4$.  } \label{phi_t}\label{fig2}
\end{figure}

The imposition of the positivity of the dark energy densities on the scalar field is also shown in FIG.-\ref{fig2}. The model parameters as constrained to ensure a positive dark energy density are $\beta=0.3$ and $n=2.5$. In this case, all five potentials provide us with a universal feature of creating a well near the bounce. The well depth may depend on the choice of the potential. The well is deeper for the case with no potential or with a generalized Starobinsky potential. However, the well is shallower for the scalar field with a constant repulsive potential.

\begin{figure}[H]
    \centering
    \includegraphics[width=72mm]{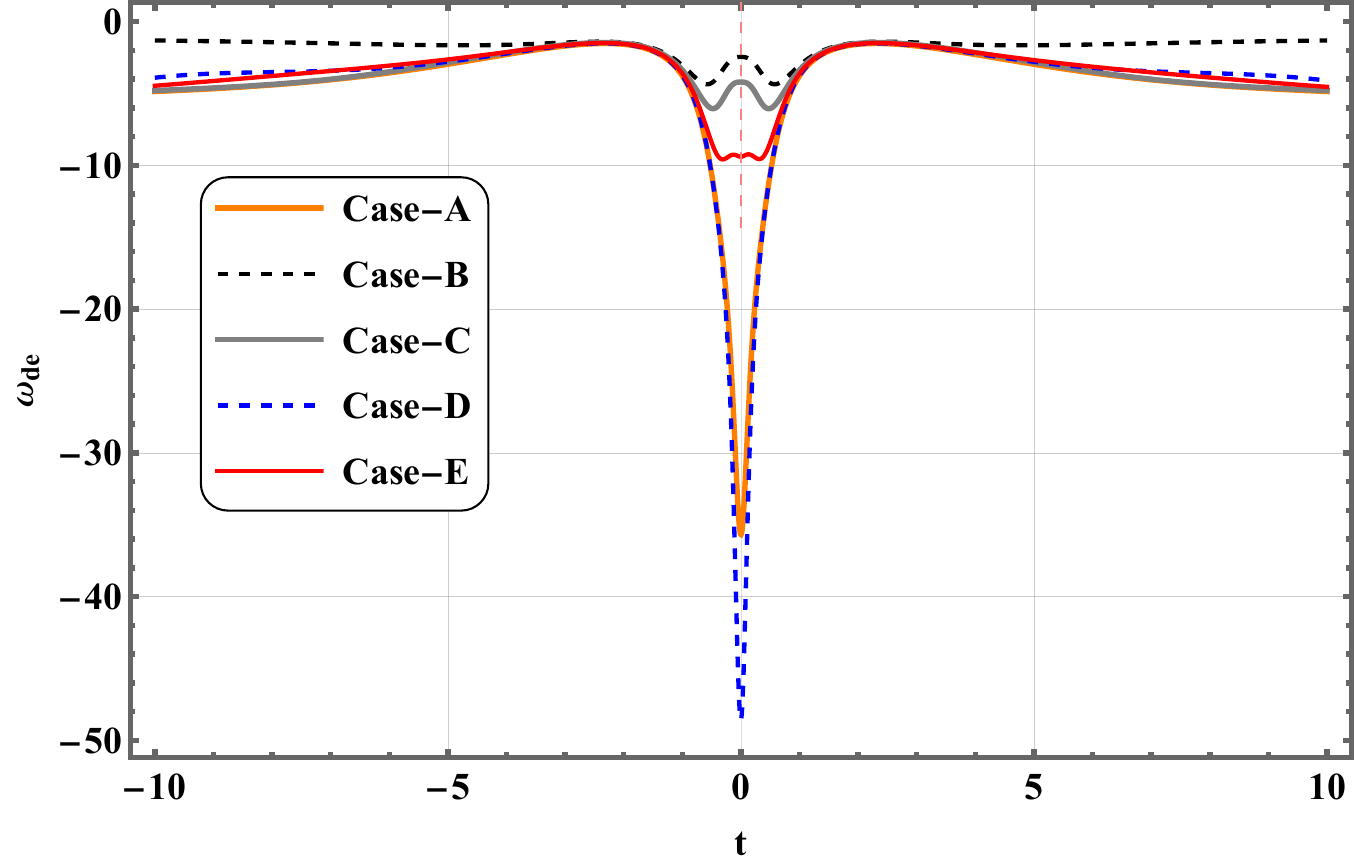}
    \caption{ Evolution of effective EoS parameter $\omega_{de}$ in cosmic time $t$ for the parametric values $\beta=-0.3, n=2.5, \gamma=0.4$. } \label{Omega_d}\label{fig3}
\end{figure}

The dynamical evolution of the Universe is shown for all the potentials through the plot of the EoS parameter in FIG.-\ref{fig3}. All of these potentials portray a gross universal feature throughout the cosmic evolution. However, near the bounce, the behaviour is potential dependent. The scalar field without interaction and with a generalized Starobinsky kind of interaction, the EoS parameter drops down near the bounce and then increases, forming a well, the depth of which is greater for the generalized Starobinsky potential case. On the other hand, the behaviour near the bounce is almost similar for all other potentials. For these later cases, the EoS parameter drops to lower values but is raised to certain upper values at the bounce.

\section{Conclusion} \label{SECV}
Teleparallel gravity theory coupled with a scalar field may be a promising theory in the context of bouncing cosmology because of their structure and effective generation of geometrical dark energy. On the other hand, scalar field theories with different inflationary potentials have proved to describe the early phase of cosmic evolution. In the present paper, we have explored some of the inflationary potentials and studied the cosmic evolutionary aspect near the bounce within the framework of $f(T,\phi)$ gravity, where we have proposed a specific form of the non-minimal coupling of the torsion and the scalar field. The scalar field and other dynamical quantities are obtained numerically after constraining the model parameters from energy conditions. We have included the case of a constant potential or a case of non-interacting scalar field for a comparison with the well-known $\alpha-$attractor potentials and the chaotic potential. More or less, all the potentials in general, provide a universal evolutionary behaviour for the scalar field and dark energy density. The evolutionary behaviour of the EoS parameter as obtained from the chaotic potential, the hyperbolic $\alpha-$attractor potential, and the constant interaction potential show similar behaviour near the bounce. On the other hand, the generalized Starobinsky potential projects similar evolutionary aspects for the dynamical quantities with the scalar field case without potential.

\section*{Acknowledgements} BM acknowledges SERB-DST for the Mathematical Research Impact Centric Support (MATRICS)[File No : MTR/2023/000371]. BM, SAK, SKT  thank IUCAA, Pune (India) for providing support in the form of an academic visit during which this work is accomplished. 
\section*{References}
\bibliography{references}

\end{document}